\documentclass[10pt]{article}

\usepackage{amsmath}
\usepackage{amssymb}

\usepackage{graphicx}

\usepackage{cite}

\usepackage{color} 

\usepackage[labelfont=bf,labelsep=period,justification=raggedright]{caption}

\makeatletter
\renewcommand{\@biblabel}[1]{\quad#1.}
\makeatother

\date{}

\begin{document}

% Title must be 150 characters or less
\begin{flushleft}
{\Large
\textbf{Pluralistic Modeling of Complex Systems}
}
% Insert Author names, affiliations and corresponding author email.
\\
Dirk Helbing$^{1,2}$%, 
%Author2$^{2}$, 
%Author3$^{3,\ast}$
\\
\bf{1} ETH Zurich, CLU, Clausiusstr. 50, 8092 Zurich, Switzerland
\\
\bf{2} Santa Fe Institute, 1399 Hyde Park Road, Santa Fe, NM 87501, USA
\\
$\ast$ E-mail: %Corresponding 
dhelbing@ethz.ch
\end{flushleft}

% Please keep the abstract between 250 and 300 words
\section*{Abstract}
The modeling of complex systems such as ecological or socio-economic systems can be very challenging. Although various modeling approaches exist, they are generally not compatible and mutually consistent, and empirical data often do not allow one to decide what model is the right one, the best one, or most appropriate one. Moreover, as the recent financial and economic crisis shows, relying on a single, idealized model can be very costly. This contribution tries to shed new light on problems that arise when complex systems are modeled. While the arguments can be transferred to many different systems, the related scientific challenges are illustrated for social, economic, and traffic systems. The contribution discusses issues that are sometimes overlooked and tries to overcome some frequent misunderstandings and controversies of the past. At the same time, it is highlighted how some long-standing scientific puzzles may be solved by considering non-linear models of heterogeneous agents with spatio-temporal interactions.
As a result of the analysis, it is concluded that a paradigm shift towards a pluralistic or possibilistic modeling approach, which integrates multiple world views, is overdue. In this connection, it is argued that it can be useful to combine many different approaches to obtain a good picture of reality, even though they may be inconsistent. Finally, it is identified what would be profitable areas of collaboration between the socio-economic, natural, and engineering sciences.

% Please keep the Author Summary between 150 and 200 words
% Use first person. PLoS ONE authors please skip this step. 
% Author Summary not valid for PLoS ONE submissions.   
%\section*{Author Summary}

\section{Introduction}

When the father of sociology, August Comte, came up with the idea of a ``social physics'', he hoped that the puzzles of social systems could be revealed with a natural science approach \cite{Comte}. However, progress along these lines was very difficult and slow. Today, most sociologists do not believe in his positivistic approach anymore. The question is whether this proves the failure of the positivistic approach or whether it just shows that social scientists did not use the right methods so far. After all, social scientists rarely have a background in the natural sciences, while the positivistic approach has been most successful in fields like physics, chemistry, or biology. 

In fact, recently new scientific communities are developing, and they are growing quickly. They call themselves socio-physicists, mathematical sociologists, computational social scientists, agent-based modelers, complexity or network scientists. Researchers from the social sciences, physics, computer science, biology, mathematics, and artificial intelligence research are addressing the challenges of social and economic systems with mathematical or computational models and lab or web experiments. Will they end up with resignation in view of the complexity of social and economic systems, or will they manage to push our knowledge of social systems considerably beyond what was imaginable even a decade ago? Will August Comte's vision of sociology as ``the queen of the sciences'' \cite{Comte2} finally become true? 

My own judgement is that it is less hopeless to develop mathematical models for social systems than most social scientists usually think, but more difficult than most natural scientists imagine. The crucial question is, how one can make substantial progress in a field as complicated and multi-faceted as the the social sciences, and how the current obstacles can be overcome? And what are these obstacles, after all? The current contribution tries to make the controversial issues better understandable to scientific communities with different approaches and backgrounds. While each of the points may be well-known to {\it some} scientists, they are probably not so obvious for others. Putting it differently, this contribution tries to build bridges between different disciplines interested in similar subjects, and make thoughts understandable to scientific communities with different points of views. 

A dialogue between social, natural and economic sciences seems to be desireable not only for the sake of an intellectual exchange on fundamental scientific problems. It also appears that science is lacking behind the pace of upcoming socio-economic problems, and that we need to become more efficient in addressing practical problems \cite{practical}. President Lee C. Bollinger of New York's prestigious Columbia University formulated the challenge as follows: ``The forces affecting societies around the world ... are powerful and novel. The spread of global market systems ... are ... reshaping our world ..., raising profound questions. These questions call for the kinds of analyses and understandings that academic institutions are uniquely capable of providing. Too many policy failures are fundamentally failures of knowledge.'' \cite{Bollinger} 

The fundamental and practical scientific challenges require from us that we do everything we can to find solutions, and that we do not give up before the limits or failure of a scientific approach have become obvious. As will be argued in Sec. \ref{final}, different methods should be seen complementary to each other and, even when inconsistent, may allow one to get a better picture than any single method can do, no matter how powerful it may seem. 

\section{Particular Difficulties of Modeling Socio-Economic Systems}

When speaking about socio-economic systems in the following, it could be anything from families over social groups or companies up to countries, markets, or the world economy including the financial system and the labor market.
The constituting system elements or system components would be individuals, groups, or companies, for example, depending on the system under consideration and the level of description one is interested in.

On the macroscopic (systemic) level, social and economic systems have some features that seem to be similar to properties of some physical or biological systems. One example is the hierarchical organization. In social systems, individuals form groups, which establish organizations, companies, parties, etc., which make up states, and these build communities of states (like the United States or the European Union, for example). In physics, elementary particles form atoms, which create molecules, which may form solid bodies, fluids or gases, which together make up our planet, which belongs to a solar system, and a galaxy. In biology, cells are composed of organelles, they form tissues and organs, which are the constituting parts of living creatures, and these make up ecosystems. 

Such analogies are certainly interesting and have been discussed, for example, by Herbert Spencer \cite{Spencer} and later on in systems theory \cite{Bertalanffy}. It is not so obvious, however, how much one can learn from them. While physical systems are often well understood by mathematical models, biological and socio-economic systems are usually not. This often inspires physicists to transfer their models to biological and socio-economic models (see the discussion in Sec. \ref{inadequate}), while biologists, social scientists, and economists often find such attempts ``physicalistic'' and inadequate. In fact, social and economic systems possess a number of properties, which distinguish them from most physical ones:
\begin{enumerate}
\item the number of variables involved is typically (much) larger (considering that each human brain contains about 1000 billion neurons), 
\item the relevant variables and parameters are often unknown and hard to measure (the existence of ``unknown unknowns'' is typical),
\item the time scales on which the variables evolve are often not well separated from each other, 
\item the statistical variation of measurements is considerable and masks laws of social behavior, where they exist (if they exist at all),
\item frequently there is no ensemble of equivalent systems, but just one realization (one human history),
\item empirical studies are limited by technical, financial, and ethical issues,
\item it is difficult or impossible to subdivide the system into simple, non-interacting subsystems that can be separately studied,
\item the observer participates in the system and modifies social reality,
\item the non-linear and/or network dependence of many variables leads to complex dynamics and structures, and sometimes paradoxical effects, 
\item interaction effects are often strong, and emergent phenomena are ubiquitous (hence, not understandable by the measurement and quantification of the individual system elements), 
\item factors such as a large degree of randomness and heterogeneity,  memory, anticipation, decision-making, communication, consciousness, and the relevance of intentions and individual interpretations complicate the analysis and modeling a lot,
\item the same applies to human features such as emotions, creativity, and innovation,
\item the impact of information is often more decisive for the behavior of a socio-economic system than physical aspects (energy, matter) or our biological heritage,
\item the ``rules of the game'' and the interactions in a social or economic system may change over time, in contrast to what we believe to be true for the fundamental laws and forces of physics,
\item in particular, social systems are influenced by normative and moral issues, which are variable.
\end{enumerate}
For such reasons, social systems are the most complex systems we know. They are certainly more complex than physical systems are. As a consequence, a considerable fraction of sociologists thinks that mathematical models for social systems are destined to fail, while most economists and many quantitatively oriented social scientists seem to believe in models with many variables. Both is in sharp contrast to the often simple models containing a few variables only that physicists tend to propose. So, who is right? The following discussion suggests that this is the wrong question. We will therefore discuss why different scientists, who apparently deal with the same research subject, come to so dramatically different conclusions. 
\par
It is clear that this situation has some undesireable side effects: Scientists belonging to different schools of thought often do not talk to each other, do not learn from each other, and probably reject each others' project proposals more frequently. It is, therefore, important to make the approach of each school understandable to the others.

\section{Modeling Approaches}

\subsection{Qualitative Descriptions}\label{Qualitative}

Many social scientists think that the fourteen challenges listed above are so serious that it is hopeless to come up with mathematical models for social systems. A common view is that all models are wrong. Thus, a widespread approach is to work out narratives, i.e. to give a qualitative (non-mathematical and non-algorithmic) description of reality that is as detailed as possible. This may be compared with a naturalist painting. 

Narratives are important, as they collect empirical evidence and create knowledge that is essential for modelers sooner or later. Good models require several steps of intellectual digestion, and the first and very essential one is to create a picture of the system one is interested in and to make sense of what is going on in it. This step is clearly indispensible. Nevertheless, the approach is sometimes criticized for reasons such as the following:
\begin{itemize}
\item Observation, description, and interpretation are difficult to separate from each other, since they are typically performed by the same brain (of a single scientist). Since these processes strongly involve the observer, it is hard or even impossible to provide an objective description of a system at this level of detail. Therefore, different scientists may analyze and interpret the system in different, subjective ways. What is an important aspect for one observer may be an irrelevant detail for another one, or may even be overlooked. In German, there is a saying that "one does not see the forest amongst all the trees'', i.e. details may hide the bigger picture or the underlying mechanisms. In the natural sciences, this problem has been partially overcome by splitting up observation, description, and interpretation into separate processes: measurements, statistical analysis, and modeling attempts. Many of these steps are supported by technical instruments, computers, and software tools to reduce the individual element and subjective influence. Obviously, this method can not be easily transferred to the study of social systems, as individuals and subjective interpretations can have important impacts on the overall system. 
\item Despite its level of detail, a narrative is often not suited to be translated into a computer program that would reproduce the phenomena depicted by it. When scientists try to do so, in many cases it turns out that the descriptions are ambiguous, i.e. not detailed enough to come up with a unique computer model. In other words, different programmers would end up with different computer models, producing different results. Therefore, Joshua Epstein claims: ``If you didn't grow it, you didn't explain it'' \cite{Epstein} (where ``grow'' stands here for ``simulate in the computer''). For example, if system elements interact in a non-linear way, i.e. effects are not proportional to causes, there are many different possibilities to specify the non-linearity: is it a parabola, an exponential dependence, a square root, a logarithm, a power law, ...? Or when a system shows partially random behavior, is it best described by additive or multiplicative noise, internal or external noise? Is it chaotic or turbulent behavior, or are the system elements just heterogeneous? It could even be a combination of several options. What differences would these various possibilities make?
\end{itemize}

\subsection{Detailed Models}\label{detailed}

In certain fields of computational social science or economics, it is common to develop computer models that grasp as many details as possible. They would try to implement all the aspects of the system under consideration, which are known to exist. In the ideal case, these facts would be properties, which have been repeatedly observed in several independent studies of the kind of system under consideration, preferably
in different areas of the world. In some sense, they would correspond to the overlapping part of many narratives. Thus, one could assume that these properties would be {\it characteristic} features of the kind of system under consideration, not just properties of a single and potentially quite particular system.  
\par
Despite it sounds logical to proceed in this way, there are several criticisms of this approach:
\begin{itemize}
\item In case of many variables, it is difficult to specify their interdependencies in the right way. (Just remember the many different possibilities to specify non-linear interactions and randomness in the system.) 
\item Some models containing many variables may have a large variety of different solutions, which may be highly dependent on the initial or boundary conditions, or the history of the system. This particularly applies to models containing non-linear interactions, which may have multiple stable solutions or non-stationary ones (such as periodic or non-periodic oscillations), or they may even show chaotic behavior. Therefore, depending on the parameter choice and the initial condition, such a model could show virtually {\it any} kind of behavior. While one may think that such a model would be a flexible world model, it would in fact be just a fit model. Moreover, it would probably not be very helpful to understand the mechanisms underlying the behavior of the system. As John von Neumann pointed out: ``With four parameters I can fit an elephant and with five I can make him wiggle his trunk.'' This wants to say that a model with many parameters can fit anything and explains nothing. This is certainly an extreme standpoint, but there is some truth in it.
\item When many variables are considered, it is hard to judge which ones are independent of each other and which ones not. If variables are mutually dependent, one effect may easily be considered twice in the model, which would lead to biased results. Dependencies among variables may also imply serious problems in the process of parameter calibration. The problem is known, for example, from sets of linear equations containing collinear variables. 
\item Models with many variables, particularly non-linear ones, may be sensitive to the exact specification of parameters, initial, or boundary conditions, or to small random effects. Phenomena like hysteresis (history-dependence) \cite{hysteresis}, phase transitions \cite{phasetrans} or ``catastrophes'' \cite{Zeeman}, chaos \cite{chaos}, or noise-induced transitions \cite{noiseind} illustrate this clearly. 
\item The parameters, initial and boundary conditions of models with many variables are hard to calibrate. If small (or no) data sets are available, the model is under-specified, and the remaining data must be estimated based on ``expert knowledge'', intuition or rules of thumb, but due to the sensitivity problem, the results may be quite misleading. The simulation of many scenarios with varying parameters can overcome the problem in part, as it gives an idea of the possible variability of systemic behaviors. However, the resulting variability can be quite large. Moreover, a full exploration of the parameter space is usually not possible when a model contains many parameters, not even with supercomputers.
\item In models with many variables, it is often difficult to identify the mechanism underlying a certain phenomenon or system behavior. The majority of variables may be irrelevant for it. However, in order to understand a phenomenon, it is essential to identify the variables and interactions (i.e. the interdependencies among them) that matter. 
\end{itemize}

\subsection{Simple Models}\label{Simple}

Simple models try to avoid (some of) the problems of detailed models by restricting themselves to a minimum number of variables needed to reproduce a certain effect, phenomenon or system behavior. They are aiming at a better understanding of so-called ``stylized facts'', i.e. simplified, abstracted, or ``ideal-typical'' observations (the ``essence''). For example, while detailed descriptions pay a lot of attention to the particular content of social norms or opinions and how they change over time in relation to the respective cultural setting, simple models abstract from the content of social norms and opinions. They try to formulate general rules of how social norms come about or how opinions change, independently of their content, with the aim of  understanding {\it why} these processes are history-dependent (``hysteretic'') and in what way they dependent on microscopic and macroscopic influences.
\par
It is clear that simple models do not describe (and do not even {\it want} to describe) all details of a system under consideration, and for this reason they are also called minimal or toy models sometimes. The approach may be represented by a few quotes. The ``KISS principle'' of model building demands to ``\underline{k}eep \underline{i}t \underline{s}imple and \underline{s}traightforward'' \cite{Kiss}. This is also known as Occam's (or Ockham's) razor, or as principle of parsimony. Albert Einstein as well demanded \cite{Einstein}: ``Make everything as simple as possible, but not simpler''. 
\par
A clear advantage of simple models is that they may facilitate an analytical treatment and, thereby, a better understanding. Moreover, it is easy to extend simple models in a way that allows one to consider a heterogeneity among the system components. This supports the consideration of effects of individuality and the creation of simple ``ecological models'' for socio-economic systems. Nevertheless, as George Box puts it: ``Essentially, all models are wrong, but some are useful'' \cite{Box}.
\par
The last quote touches an important point. The choice of the model and its degree of detail should depend on the purpose of a model, i.e. its range of application. For example, there is a large variety of models used for the modeling and simulation of freeway traffic. The most prominent model classes are ``microscopic'' car-following models, focussing on the interaction of single vehicles, ``mesoscopic'' gas-kinetic models, describing the change of the velocity distribution of vehicles in space and time, ``macroscopic'' fluid-dynamic models, restricting themselves to changes of the average speed and density of vehicles, and cellular automata, which simplify microscopic ones in favor of simulation speed. Each type of model has certain ranges of application. Macroscopic and cellular automata models, for example, are used for large-scale traffic simulations to determine the traffic situation on freeways and perform short-term forecasts, while microscopic ones are used to study the interaction of vehicles and to develop driver assistance systems. For some of these models, it is also known how they are mathematically connected with each other, i.e. macroscopic ones can be derived from microscopic ones by certain kinds of simplifications (approximations) \cite{EPJB1,RMP}.
\par
The main purpose of models is to guide people's thoughts. Therefore, models may be compared with city maps. It is clear that maps simplify facts, otherwise they would be quite confusing. We do not want to see any single detail (e.g. each tree) in them. Rather we expect a map to show the facts we are interested in, and depending on the respective purpose, there are quite different maps (showing streets, points of interest, topography, supply networks, industrial production, mining of natural resources, etc.). 
\par
One common purpose of models is {\it prediction}, which is mostly (mis)understood as ``forecast'', while it often means ``the identification of implications regarding how a system is expected to behave under certain conditions''. It is clear that, in contrast to the motion of a planet around the sun, the behavior of an individual can hardly be forecasted. Nevertheless, there are certain tendencies or probabilities of doing certain things, and we usually have our hypotheses of what our friends, colleagues, or family members would do in certain situations. It turns out that, when many people interact, the aggregate behavior can sometimes be quite predictable. For example, the ``wisdom of crowds'' is based on the statistical law of large numbers \cite{Wisdom}, according to which individual variations (here: the independent estimation of facts) are averaged out. 
\par
Furthermore, interactions between many individuals tend to restrict the degree of freedom regarding what each individual can or will do. This is, why the concept of ``social norms'' is so important. Another example is the behavior of a driver, which is constrained by other surrounding vehicles. Therefore, the dynamics of traffic flows can be mathematically well understood \cite{RMP,analytical}. Nevertheless, one cannot exactly forecast the {\it moment} in which free traffic flow breaks down and congestion sets in, and therefore, one cannot forecast travel times well. The reason for this is the history-dependent dynamics, which makes it dependent on random effects, namely on the size of perturbations in the traffic flow. However, what can be predicted is what are the possible traffic states and what are conditions under which they can occur. One can also identify the probability of traffic flows to break down under certain flow conditions, and it is possible to estimate travel times under free and congested flow conditions, given a measurement of the inflows. The detail that cannot be forecasted is the exact moment in which the regime shift from free to congested traffic flow occurs, but this detail has a dramatic influence on the system. It can determine whether the travel time for a certain freeway section is 2 minutes or 20 minutes.
\par
However, it is important to underline that, in contrast to what is frequently stated, the purpose of developing models is not only prediction. Joshua Epstein, for example, discusses 16 other reasons to build models, including explanation, guiding data collection, revealing dynamical analogies, discovering new questions, illuminating core uncertainties, demonstrating tradeoffs, training practitioners, and decision support, particularly in crises \cite{Epstein2}. 
\par
Of course, not everybody favors simple models, and typical criticisms of them are:
\begin{itemize}
\item It is usually easy to find empirical evidence, which is not compatible with simple models (even though, to be fair, one would have to consider the purpose they have been created for, when judging them). Therefore, one can say that simple models tend to over-simplify things and leave out more or less important facts. For this reason, they may be considered inadequate to describe a system under consideration.
\item Due to their simplicity, it may be dangerous to take decisions based on their implications.
\item It may be difficult to decide, what are the few relevant variables and parameters that a simple model should consider. Scientists may even disagree about the stylized facts to model.
\item Simple models tend to reproduce a few stylized facts only and are often not able to consistently reproduce a large number of observations. The bigger picture and the systemic view may get lost.
\item Making simple models compatible with a long list of stylized facts often requires to improve or extend the models by additional terms or parameter dependencies. Eventually, this improvement process ends up with detailed models, leaving one with the problems specified there (see Sec. \ref{detailed}).
\item Certain properties and behaviors of socio-economic systems may not be understandable with methods that have been successful in physics: Subdividing the system into subsystems, analyzing and modeling these subsystems, and putting the models together may not lead to a good description of the overall system. For example, several effects may act in parallel and have non-separable orders of magnitude. 
This makes it difficult or impossible to start with a zeroth or first order approximation and to improve it by adding correction terms (as it is done, for example, when the falling of a body is described by the effect of gravitational acceleration plus the effect of air resistance). Summing up the mathematical terms that describe the different effects may not converge. It is also not clear whether complex systems can be always understood via simple principles, as the success of complexity science might suggest. Some complex systems may require {\it complex} models to explain them, and there may even be phenomena, the complexity of which is irreducible. Turbulence \cite{turbulence} could be such an example. While it is a long-standing problem that has been addressed by many bright people, it has still not been explained completely. 
\end{itemize}
It should be added, however, that we do not know today, whether the last point is relevant, how relevant it is, and where. So far, it is a {\it potential} problem one should be aware of. It basically limits the realm, in which classical modeling will be successful, but we have certainly not reached these limits, yet. 

\subsection{Modeling Complex Systems}

Modeling socio-economic systems is less hopeless than many social scientists may think \cite{Weidlich}. In recent years, considerable progress has been made in a variety of relevant fields, including
\begin{itemize}
\item experimental research \cite{exp1,exp2,FutureExperimenting},
\item data mining \cite{datamining},
\item network analysis \cite{networks},
\item agent-based modeling \cite{Epstein,agentbased},
\item the theory of complex systems (including self-organization phenomena and chaos) \cite{complex},
\item the theory of phase transitions \cite{phasetrans} (``catastrophes'' \cite{Zeeman}), critical phenomena \cite{criticalphen}, and extreme events \cite{extreme}, and
\item the engineering of intelligent systems \cite{AI,robotics}.
\end{itemize}
These fields have considerably advanced our understanding of complex systems. In this connection, one should be aware that the term ``complexity'' is used in many different ways. In the following, we will distinguish three kinds of complexity: 
\begin{enumerate}
\item structural, 
\item dynamical, and 
\item functional complexity. 
\end{enumerate}
One could also add algorithmic complexity, which is given by the amount of computational time needed
to solve certain problems. Some optimization problems, such as the optimization of logistic or traffic signal operations, are
algorithmically complex \cite{BioLogistics}.
\par
Linear models are not considered to be complex, no matter how many terms they contain. An example for {\it structural} complexity is a car or airplane. They are constructed in a way that is dynamically more or less deterministic and well controllable, i.e. dynamically simple, and 
they also serve relatively simple functions (the motion from a location A to another location B). While the acceleration of a car or a periodic oscillation would be an example for a simple {\it dynamics}, examples for complex dynamical behavior are non-periodic
changes, deterministic chaos, or history-dependent behaviors. Complex dynamics can already be
produced by simple sets of non-linearly coupled equations. While a planet orbiting around the sun follows
a simple dynamics, the interaction of three celestial bodies can already show a chaotic dynamics. Ecosystems,
the human body or the brain would be {\it functionally} complex systems. The same would hold for the world wide web, 
financial markets, or running a country or multi-national company. 
\par
While the interrelation between function, form and dynamics 
still poses great scientific challenges, the understanding of structurally or dynamically complex systems has significantly progressed. Simple agent-based models of systems with a large number of interacting system elements (be it particles, cars, pedestrians, 
individuals, or companies) show properties, which remind of socio-economic systems. Assuming that these
elements mutually adapt to each other through non-linear or network interactions (i.e. that the elements are influenced by their environment while modifying it themselves), one can find a rich, history-dependent system behavior, which is often counter-intuitive, hardly predictable, and seemingly uncontrollable. These models challenge our common way of thinking and help to grasp behaviors of complex systems, which are currently a nightmare for decision-makers. 
\par
For example, complex systems are often unresponsive to control attempts, while close to ``critical points'' (also known as ``tipping points''), they may cause sudden (and often unexpected) phase transition (so-called ``regime shifts''). These correspond to discontinuous changes in the system behavior. The breakdown of free traffic flow would be a harmless example, while a systemic crisis (such as a financial collapse or revolution) would be a more dramatic one. Such systemic crises are often based on cascade spreading through network interactions \cite{SystemicInstability}. Complex adaptive systems also allow one to understand extreme events as a result of strong interactions in a system (rather than as externally caused shocks). Furthermore, the interaction of many system elements may give rise to interesting self-organization phenomena and emergent properties, which cannot be understood from the behaviors of the single elements or by adding them up. Typical examples are collective patterns of motion in pedestrian crowds or what is sometimes called ``swarm intelligence'' \cite{TopiCS}.
\par
Considering this, it is conceivable that many of today's puzzles in the social sciences may one day be explained by simple models, namely as {\it emergent} phenomena resulting from interactions of many individuals and/or other system elements. Note that emergent phenomena cannot be explained by {\it linear} models (which are most common in many areas of quantitative empirical research in the social sciences and economics). Unfortunately, there is no standard way to set up models of emergent phenomena. On the one hand, there are many possible kinds of non-linear functional dependencies (``interactions'') (see the end of Sec. \ref{Qualitative}). On the other hand, model assumptions that appear plausible do often not produce the desired or expected effects. 
\par
In spite of these difficulties, taking into account time-dependent change, a non-linear coupling of variables, spatial or network interactions, randomness, and/or correlations (i.e. features that many social and economic models currently do not consider to the necessary extent), can sometimes deliver unexpected solutions of long-standing puzzles. For example, it turns out that representative agent models (which are common in economics) can be quite misleading, as the same kinds of interactions among the system components can imply completely different or even opposite conclusions, when interactions take place in a socio-economic network rather than with average (or randomly chosen) interaction partners \cite{PLoS}. In other words, models often produce counter-intuitive results, when spatio-temporal or network interactions are relevant. Therefore, a simple non-linear model may explain phenomena, which complicated linear models may fail to reproduce. In fact, this generally applies to systems that can show several possible states (i.e. systems which do not have just one stable equilibrium). Out-of-equilibrium models are also required for the description of systemic crises such as the current financial crisis \cite{SystemicInstability}.  

\section{Challenges of Socio-Economic Modeling}

Many people before and after Popper have been thinking about the logic of scientific discovery \cite{Popper}. A wide-spread opinion is that a good model should be applicable to measurements of many systems of a certain kind, in particular to measurements in different areas of the world. The more observations a model can explain and the less parameters it has, the more powerful it is usually considered to be. Models with a {\it few} parameters can often be easier to calibrate, and cause-and-effect relationships may be better identified, but one can usually not expect that
these models would provide an {\it exact} description of reality. Nevertheless, a good model should make predictions regarding some possible, but previously unobserved system behaviors. In this connnection, prediction does not necessarily mean the {\it forecast} of a certain event at a specific future point in time. It means a specific system behavior that is expected to occur (or to be possible) under certain conditions (e.g. for certain parameter combinations or certain initial conditions). When such conditions apply and the system shows the expected behavior, this would be considered to verify the model, while the model would be falsified or seriously questioned, if the predicted system behavior would not occur. By experimentally challenging models based on their predictions (implications), it has been possible in the natural sciences to rate alternative models based on their quality in reproducing and predicting measured data. Unfortunately, it turns out that this approach is less suited to identify ``the right model'' of a social or economic system under consideration. As we will discuss in the following, this is not only due to the smaller amount of data available on most aspects of social and economic systems and due to experimental limitations for financial, technical and ethical reasons... 

\subsection{Promises and Difficulties of the Experimental Approach}

So far, it is very expensive to carry out social and economic experiments, for example in the laboratory. While
the study of human behavior under controlled conditions has become a common research method not only in
psychology, but also in experimental economics and in sociology, the number of individuals that can be studied
in such experiments is limited. This implies a large degree of statistical variation, which makes it difficult to determine
behavioral laws or to distinguish between different models. The statistical noise creates something like a foggy 
situation, which makes it difficult to see what is going on. In physics, this problem can be usually solved
by better measurement methods (apart from uncertainty that results from the laws of quantum mechanics).
In social systems, however, there is an irreducible degree of randomness. The behavior varies not only between individuals
due to their heterogeneity (different ``personality''). It also varies from one instance to another, i.e. 
the decision-making of an individual is usually not deterministic. This could be due to various reasons: unknown
external influences (details attracting the attention of the individual) or internal factors (exploration behavior,
decisions taken by mistake, memory effects, etc.). The large level of behavioral variability within and between
individuals is probably not only due to the different histories individuals have, but also due to the fact that exploration
behavior and the heterogeneity of behaviors are beneficial for the learning of individuals and for the adaptibility of 
human groups to various environmental conditions. Applying a theory of social evolution would, therefore, suggest
that randomness is significant in social and economic systems, because it increases system performance. Besides,
heterogeneity can also have {\it individual} benefits, as differentiation facilitates specialization. The benefit of a variation {\it between} individuals is also well-known from ecological systems \cite{inLevin}.
\par
\begin{figure}[htbp]
\begin{center}
\includegraphics[width=12cm]{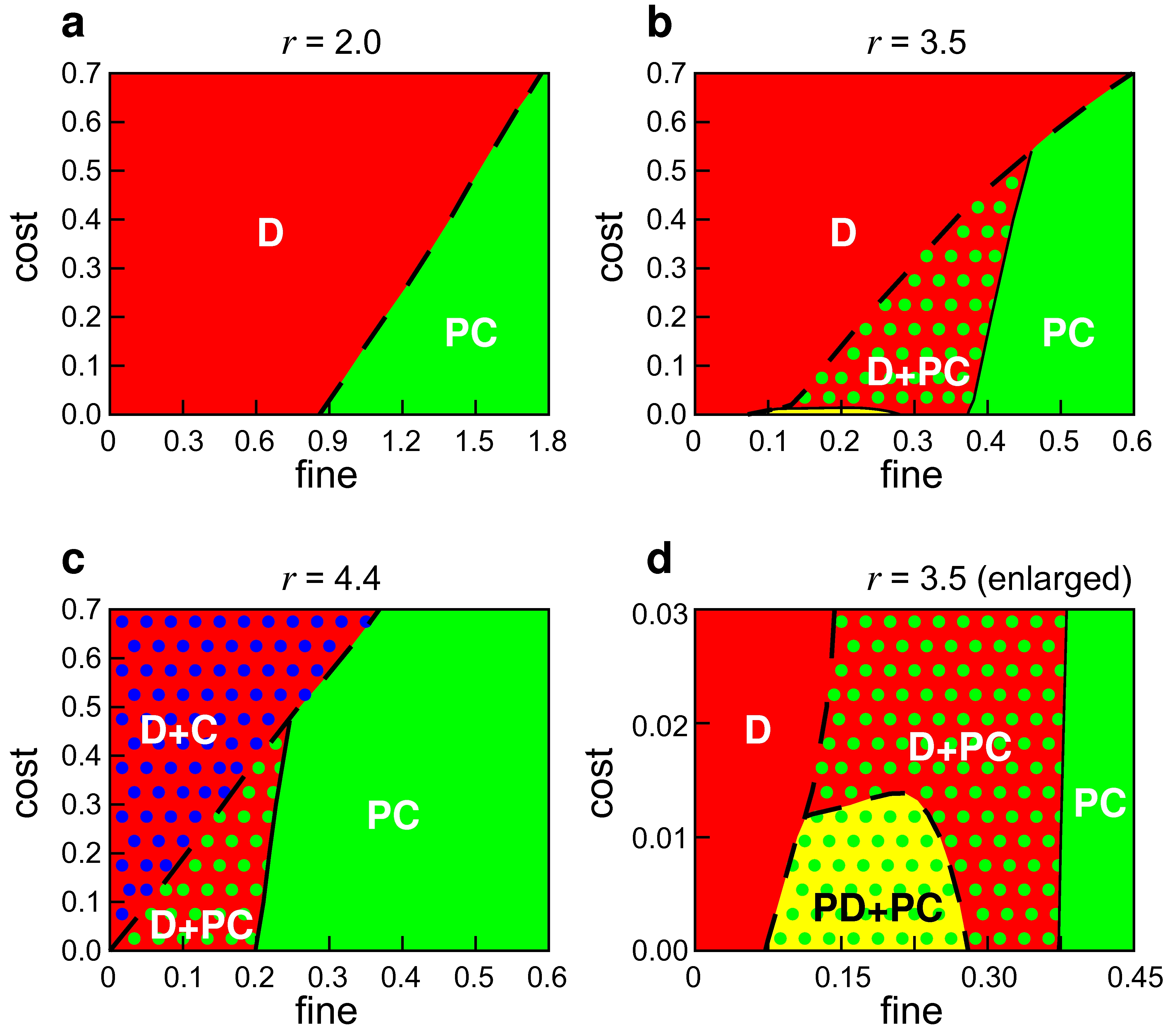}
\end{center}
\caption{Phase diagram showing the finally remaining strategies in the spatial public goods game with cooperators (C), defectors (D), cooperators who punish defectors (PC) and hypocritical punishers (PD), who punish other defectors while defecting themselves (after Ref. \cite{PLoS}). Initially, each of the four strategies occupies 25\% of the sites of the square lattice, and their distribution is uniform in space. However, due to their evolutionary competition, two or three strategies die out after some time. The finally resulting state depends on the punishment cost, the punishment fine, and the synergy $r$ of cooperation (the factor by which cooperation increases the sum of investments). The displayed phase diagrams are for (a) $r=2.0$, (b) $r=3.5$, and (c) $r=4.4$. (d) Enlargement of the small- cost area for $r=3.5$. Solid separating lines indicate that the resulting fractions of all strategies change continuously with a modification of the punishment cost and punishment fine, while broken lines correspond to discontinuous changes. All diagrams show that cooperators and defectors cannot stop the spreading of costly punishment, if only the fine-to-cost ratio is large enough (see green PC area). Note that, in the absence of defectors, the spreading of punishing cooperators is extremely slow and follows a voter model kind of dynamics. A small level of strategy mutations (which continuously creates a small number of strategies of all kinds, in particular defectors) can largely accelerate the spreading of them. Furthermore, there are parameter regions where punishing cooperators can crowd out "second-order free-riders'' (non-punishing cooperators) in the presence of defectors (D+PC). Finally, for low punishment costs, but moderate punishment fines, it may happen that "moralists'', who cooperate and punish non-cooperative behavior, can only survive through an "unholy alliance'' with "immoral'', hypocritical punishers (PD+PC). For related videos, see http://www.soms.ethz.ch/research/secondorder-freeriders  or http://www.matjazperc.com/games/moral.html.}\label{Fig1}
\end{figure}

Besides impeding the discovery of behavioral laws, the limited number of participants in laboratory experiments
also restricts the number of repetitions and the number of experimental settings or parameter combinations that can be studied. Scanning parameter spaces is impossible so far, while it would be useful to detect different system behaviors and to determine under which conditions they occur. It can be quite tricky to select suitable system parameters (e.g. the payoff matrix in a game-theoretical experiment). Computer simulations suggest that one would find interesting results mainly, if the parameters selected in different experimental setups imply different system behaviors, i.e. if they belong to different ``phases'' in the parameter space (see Fig. \ref{Fig1}). In order to determine such parameter combinations, it is advised to perform computer simulations {\it before}, to determine the phase diagram for the system under consideration \cite{FutureExperimenting}. The problem, however, is that the underlying model is unlikely to be perfect, i.e. even a good social or economic model is expected to make predictions which are only approximately valid. As a consequence, the effect one likes to show may appear for (somewhat) different parameter values, or it may not occur at all (considering the level of randomness) \cite{TraulsenExp}. 
\par

\subsection{Several Models Are Right}

The above mentioned properties of socio-economic systems imply that it is difficult to select the ``right'' model among several alternative ones. For an illustration, let us take car-following models, as they are used for the simulation of urban or freeway traffic. Thanks to radar sensors, it has become possible to measure the acceleration of vehicles as a function of the typical variables of car-following models, which are the distance to the car ahead, the own speed, and the speed difference. When fitting the parameters of different car-following models to data of such measurements, it turns out that the remaining error between computer simulations and measurements is about the same for most of the models. The calibration error varies between 12 and 17 percent, and according to the authors, ``no model can be denoted to be the best'' \cite{Wagner}. When the error of different models (i.e. the deviation between model and data) is determined for a {\it new} data set (using the model parameters determined with the {\it previous} data set), the resulting validation error usually varies between 17 and 22 percent  (larger validation errors mainly result, when the calibration data set is overfitted) \cite{Wagner}. Again, the performance of the different models is so similar that it would not be well justified to select one of them as the ``correct'' model and exclude all the others. A closer analysis shows that the parameters of the car-following dynamics varies among different drivers, but the behavior of specific drivers also vary over time \cite{variable}. We have to assume that the same applies to basically all kinds of behavior, not only for driving a car. Moreover, it is likely that many behaviors (such as decision-making behaviors) vary even more than car-following behavior does. As a consequence, it would be even more difficult to distinguish between different models by means of empirical or experimental data, which would mean that we may have to accept several models to be (possibly) ``right'', even when they are not consistent with each other. In other words, the question ``What is the best model?'' or ``How to choose the model?'' may not be decidable in a reasonable way, as is also suggested by the next section. This situation reminds a bit of G\"odel's Undecidability Theorem \cite{Goedel}, 
which relates to the (in)completeness of certain axiom systems. 
\par
It may be tempting to determine the best model as the one which is most {\it successful}, for example in terms of the number of citations it gets. However, success is not necessarily an indicator of a good model. Let us take models used for stock trading as an example. Clearly, even if the stock prices vary in a perfectly random manner and if the average success of each model is the same over an infinite time period; when different traders apply different trading models, they will be differently successful at any chosen point in time. Therefore, one would consider some models more successful than others, while this would be only a matter of luck. As a matter of chance, at other points in time, different models would be the most successful ones. 
\par
Of course, if behaviors are not just random so that behavioral laws that go beyond statistical distributions exist, some models should be better than others, and it should eventually be possible to separate ``good'' from ``bad'' models through the ``wisdom of crowds'' effect. However, the ``wisdom of crowds'' assumes independent judgements, while scientists have repeated interactions. It has be shown experimentally that this tends to create consensus, but that this consensus will often deviate from the truth \cite{WisdomExp}. The problem results from social influence, which creates a herding effect that can undermine the ``wisdom of crowds''. Of course, this mainly applies, when the facts are not sufficiently obvious, which is the case in the social sciences due to the high variability of observations, while the problem is less pressing in the natural sciences thanks to the higher measurement precision. Nevertheless, the physicist Max Planck is known for the quote: ``Science progresses funeral by funeral'' \cite{Planck}. 
Kuhn's study of scientific revolutions \cite{Kuhn} suggests as well that scientific progress is not continuous, but there are sudden paradigm shifts. This reveals the problem of herding effects. Even a collective agreement is no guarantee for the correctness of a model, as the replacement of classical mechanics by relativistic quantum theory shows. In other words, success is no necessarily an indicator for good models. It may just be an indicator for what model is most fashionable at a given time. The problem becomes worse by the academic selection process that decides, what scientists make a carreer and which ones not. It creates a considerable inertia in the adjustment to new knowledge, i.e. scientific trends are likely to persist longer than what is justified by facts. 

\subsection{No Known Model is Right}

A typical approach in the natural sciences is to verify or falsify previously untested predictions (implications) of alternative models by sometimes quite sophisticated experiments. Only in the minority of cases, two alternative theories turn out to be the same, like the wave and the particle picture of quantum mechanics. In most cases, however, two theories A and B are non-identical and inconsistent, which means that they should make different predictions in particular kinds of situations. Experiments are performed to find out whether theory A or theory B is right, or whether both of them deviate from the measurements. If the experimental data confirm theory A and are not compatible with theory B (i.e. deviate significantly from it), one would discard theory B forever. In this way, experiments are thought to narrow down the number of alternative theories, until a single, ``true'' theory remains.
\par
When social or economic systems are modeled, the following situation is not unlikely to happen: Scientists identify mutually incompatible predictions of theories A and B, and it turns out that an experiment supports theory A, but not theory B. One day, another scientist identifies a different set of incompatible predictions, and another experiment supports theory B, but not theory A. Due to the inherent simplifications of socio-economic models, for {\it any} model it should easy to find empirical evidence that contradicts it. What should one do in such cases? Giving up on modeling would probably not be the best idea. Generalizing a model is always possible, but it will usually end up with detailed models, which implies a number of problems that have been outlined in Sec. \ref{...}. One could also stay with many particular models and determine their respective ranges of validity. This, however, will never result in a holistic or systemic model. A possible way out would be the approach of pluralistic modeling outlined in Sec. \ref{...}.
\par
Modeling in modern physics seems to face similar problems. While one would expect that each experiment narrows
down the number of remaining, non-falsified models, one actually observes that, after each experiment, scientists come up with
a number of new models. As people say: "Each answered question raises ten new ones." In fact, there is an abundance of elementary particle models, and the same applies to cosmological models. Many models require to assume the existence of factors that have never been measured and perhaps {\it will} never be measured, such as Higgs bosons, dark matter, or dark energy. We will probably have to live with the fact that models are just {\it models} that never grasp all details of reality. Moreover, as has been pointed out, understanding elementary particles and fundamental forces in physics would not explain at all what is happening in the world around us \cite{Vicsek,Pietronero}. Many emergent phenomena that we observe in the biological, economic and social world will never be derived from elementary particle physics, because emergent properties of a system cannot be understood from the properties of its system components alone. They usually come about by the interaction of a large number of system components. Let us be honest: We do not even understand the particular properties of water, as simple as H$_2$O molecules may be.
\par
Generally, there is a serious lack in understanding the connection between function, dynamics, and form. Emergence often seems to have an element of surprise. The medical effect of a new chemical drug cannot be understood by computer simulation alone. So far, we also do not understand emotions and consciousness, and we cannot calculate the biological fitness of a species in the computer. The most exciting open puzzles in science concern such emergent phenomena. It would be interesting to study, whether social and economic phenomena such as trust, solidarity, and economic value can be understood as emergent phenomena as well \cite{practical}.

\subsection{The Model Captures Some Features, But May Be Inadequate}\label{inadequate}

Scientists are often prompted to transfer their methods to another areas of application, based on analogies that they see between the behavior of different systems. Systems science is based on such analogies, and physicists generalize their methods as well. The question is how useful a ``physicalist approach'' can be, which transfers properties of many-particle systems to social or economic systems, although individuals are certainly more intelligent than particles and have many more behavioral degrees of freedom. 
\par
Of course, physicists would never claim that particle models could provide an exact description of social or economic systems. Why, then, do they think the models could make a contribution to the understanding of these systems? This is, because they have experience with what can happen in systems characterized by the non-linear interaction of many system components in space and time, and when randomness plays a role. The know how self-organized collective phenomena on the ``macroscopic'' (aggregate) level can results from interactions on the ``microscopic'' (individual) level. And they have learned, how this can lead to phase transitions (also called ``regime shifts'' or ``catastrophes''), when a system parameter (``control parameter'') crosses a critical point (``tipping point''). Furthermore, they have discovered that, at a critical point, the system typically shows a scale-free behavior (i.e. power laws or other fat-tail distributions rather than Gaussian distributions). 
\par
It is important to note, that the characteristic features of the system at the critical point tend to be {\it ``universal''}, i.e. they do not depend on the details of the interactions. {\it This} is, why physicists think they can abstract from the details. Of course, details {\it are} expected to be relevant when the system is {\it not} close to a critical point. It should also be added, that there are a couple of different kinds of universal behavior, so-called universality classes. Nevertheless, many-particle models may allow one to get a better understanding of regime shifts, which are not so well understood by most established models in economics or the social sciences. However, if the tipping point is far away, the usefulness of many-particle models is limited, and detailed descriptions, as they are favored by economists and social scientists, appear to be more adequate. 
\par
Sometimes, it is not so clear how far analogies can carry, or if they are useful at all. Let us take neural network models. In a certain sense, they can be used to model learning, generalization, and abstraction. However, the hope that they would explain the functioning of the brain has been largely disappointed. Today, we know that the brain works quite differently, but neural network theory has given birth to many interesting engineering applications that are even commerically applied. Let us consider models of cooperation based on coupled oscillators as a second example. Without any doubt, the synchronization of cyclical behavior is among the most interesting collective phenomena we know of, and models allow one to study if and how groups of oscillators will coordinate each other or fall apart into subgroups (which are not synchronized among each other, while the oscillators in each of them are) \cite{Mikhailov}. Despite this analogy to group formation and group dynamics, it is not clear, what we can learn from such models for social systems. A similar point is sometimes raised for spin models, which have been proposed to describe opinion formation processes or the emergence of cooperation in social dilemma situations. In this connection, it has been pointed out that social interactions cannot always be broken down into binary interactions. Some interactions involve three or more individuals at the same time, which may change the character of the interaction. Nevertheless, similar phenomena have been studied by overlaying binary interactions, and it is not fully clear how important the difference is. 
\par
Let us finally ask whether unrealistic assumptions are generally a sign of bad models? The discussion in Sec. \ref{Simple} suggests that this is not necessarily so. It seems more a matter of the purpose of a model, which determines the level of simplification, and a matter of the availability of better models, i.e. a matter of competition. Note, however, that a more realistic model is not necessarily more useful. For example, many car-following models are more realistic than fluid-dynamic traffic models, but they are not suited to simulate large-scale traffic networks in real-time. For social systems, there are a number of different modeling approaches as well, including the following:
\begin{itemize}
\item {\it Physical(istic) modeling approach:} Socio- and econo-physicists often abstract social interactions so much that their models come down to multi-particle models (or even spin models with two behavioral options). Such models focus on the effect of non-linear interactions and are a special case of bounded rationality models, sometimes called zero-intelligence models \cite{Omerod}. Nevertheless, they may display features of {\it collective} or {\it swarm} intelligence \cite{TopiCS}. Furthermore, they may be suited to describe regime shifts or situations of routine choice \cite{Gintis}, i.e. situations where individuals react to their environment in a more or less subconscious and automatic way. Paul Omerod, an economist by background, argues as follows \cite{Omerod2}: ``In many social and economic contexts, self-awareness of agents is of little consequence... No matter how advanced the cognitive abilities of agents in abstract intellectual terms, it is as if they operate with relatively low cognitive ability within the system... The more useful ?null model? in social science agent modelling is one close to zero intelligence. It is only when this fails that more advanced cognition of agents should be considered.''
\item {\it Economic modeling approach:} Economists seem to have quite the opposite approach. Their concept of ``homo economicus'' (the ``perfect egoist'') assumes that individuals take strategic decisions, choosing the optimal of their behavioral options. This requires individuals with infinite memory and processing capacities. Insofar, one could speak of an infinite-intelligence approach. It is also known as rational choice approach and has the advantage that the expected behaviors of individuals can be axiomatically derived. In this way, it was possible to build the voluminous and impressive theory of mainstream economics. Again, the reliability of this theory depends, of course, on the realism of its underlying assumptions.
\item {\it Sociological modeling approach:} Certain schools of sociologists use rational choice models as well. In contrast to economists, however, they do not generally assume that individuals would radically optimize their own profit. Their models rather consider that social exchange is more differentiated and multi-faceted. For example, when choosing their behavior, individuals may not only consider their own preferences, but the preferences of their interaction partner(s) as well. In recent years, ``fairness theory'' has received a particular attention \cite{Fehr} and often been contrasted with rational choice theory. These social aspects of decision-making are now eventually entering economic thinking as well \cite{Frey}.
\item {\it Psychological modeling approach:} Psychologists are perhaps least axiomatic and usually oriented at empirical observations. They have identified behavioral paradoxies, which are inconsistent with rational choice theory, at least its classical variant. For example, it turns out that most people behave in a risk averse way. To account for their observations, new concepts have been developed, including prospect theory \cite{prospect}, satisficing theory \cite{satisficing}, and the concept of behavioral heuristics \cite{Gigerenzer}. In particular, it turns out that individual decisions depend on the respective framing. In his Nobel economics lecture, Daniel Kahneman put it this way: ``Rational models are psychologically unrealistic... the central characteristic of agents is not that they reason poorly, but that they often act intuitively. And the behavior of these agents is not guided by what they are able to comppute, but by what they happen to see at a given moment.'' Therefore, modern research directions relate to the cognitive and neurosciences. These results are now finding their way into economics via the fields of experimental, behavioral, and neuro-economics. 
\end{itemize}
In summary, there is currently no unified approach that scientists generally agree on. Some of the approaches are more stylized or axiomatic. Others are in better quantitative agreement with empirical or experimental evidence, but mathematically less elaborated. Therefore, they are theoretically less suited to derive implications for the behavior in situations, which have not been explored so far. Consequently, all models have their strengths and weaknesses, no matter how realistic they may be. Moreover, none of the mathematical models available so far seems to be sophisticated enough to reflect the full complexity of social interactions between many people.

\subsubsection{Different Interpretations of the Same Model}

A further difficulty of modeling socio-economic systems is that scientists may not agree on the interpretation of a model. Let us discuss, for example, the multi-nomial logit model, which has been used to model decision-making in a large variety of contexts and awarded with the nobel prize \cite{McFadden}. This model can be derived in a utility-maximizing framework, assuming perfectly rational agents deciding under conditions of uncertainty. The very same model, however, can also be derived in other ways. For example, it can be linked to psychological laws or to distributions of statistical physics \cite{QuantSoc}. In the first case, the interpretation is compatible with the infinite-intelligence approach, while in the last case, it is compatible with the zero-intelligence approach, which is quite puzzling. A comparison of these approaches is provided by Ref. \cite{QuantSoc}.

\section{Discussion and Outlook}\label{final}

\subsection{Pluralistic or Possibilistic Modeling and Multiple World Views: The Way Out?}

Summarizing the previous discussion, it is quite unlikely that we will ever have a single, consistent, complete, and correct model of socio-economic systems. Maybe we will not even find such a grand unified theory in physics. Recently, doubts along these lines have even been raised by some particle physicists \cite{particle1,particle2}. It may be the time to say good bye to a modeling approach that believes in the feasibility of a unique, general, integrated and consistent model. At least there is no theoretical or empirical evidence for the possibility of it. 
\par
This calls for a paradigm shift in the modeling approach. It is important to be honest that each model is limited, but most models are useful for something. In other words, we should be tolerant with regard to each others models and see where they can complement each other. This does not mean that there would be separate models for non-overlapping parts of the system, one for each subsystem. As has been pointed out, it is hard to decide whether a particular model is valid, no matter how small the subsystem is chosen. It makes more sense to assume that each model has a certain validity or usefulness between 0 and 1, and that the validity furthermore depends on the part or aspect of the system addressed. This validity may be quantified, for example, by the goodness of fit of a given system or the accuracy of description of another system of the same kind. As there are often {\it several} models for each part or aspect of a system, one could weight the models with their respective validity, as determined statistically. Analogously to the ``wisdom of crowds'' \cite{Wisdom}, which is based on the law of large numbers, this should lead to a better quantitative fit or prediction than most (or even each) model in separation, despite the likely inconsistency among the models. Such an approach could be called a {\it pluralistic} modeling approach \cite{pluralistic}, as it tolerates and integrates multiple world views. It may also be called a {\it possibilistic} approach \cite{possibilistic}, because it takes into account that each model has only a certain likelihood to be valid, i.e. each model describes a possible truth. However, this should not be misunderstood as an appeal for a subjectivistic approach. The pluralistic modeling approach still assumes that there is some underlying reality that some, many, or all of us share (depending on the aspect we talk about).
\par
As shocking as it may be for many scientists and decision-makers to abandon their belief in the existence of a unique, true model, the pluralistic modeling approach is already being used. Hurricane prediction and climate modeling are such examples \cite{climate}. Even modern airplanes are controlled by multiple computer programs that are run in parallel. If they do not agree with each other, a majority decision is taken and implemented. Although this seems pretty scary, this approach has worked surprisingly well so far. Moreover, when crash tests of newly developed cars are simulated in the computer, the simulations are again performed with {\it several} models, each of which is based on different approximation methods. It is plausible to assume that pluralistic modeling will be much more widely used in future, whenever a complex system shall be modeled.

\subsection{Where Social Scientists and Natural Scientists or Engineers Can Learn From Each Other}

It has been argued that each modeling approach has its strength and weaknesses, and that they should be considered complementary rather than competitive. This also implies that scientists of different disciplines may profit and learn from each other. Areas of fruitful multi-disciplinary collaboration could be:
\begin{itemize}
\item the modeling of socio-economic systems themselves, 
\item understanding the impacts that engineered systems have on the socio-economic world, 
\item the modeling of the social mechanisms that drive the evolution and spreading of innovations, norms, technologies, products etc., 
\item scientific challenges relating to the managing of complexity and to systems design,
\item the application of social coordination and cooperation mechanisms to the creation of self-organizing technical systems (such as decentralized traffic controls or peer-to-peer systems), 
\item the development of techno-social systems \cite{Vespignani}, in which the use of technology is combined  with social competence and human knowledge (such as Wikipedia, prediction markets, recommender systems, or the semantic web). 
\end{itemize}
Given the large potentials of such collaborations, it is time to overcome disciplinary boundaries. They seem to make less and less sense. It rather appears that multi-disciplinary, large-scale efforts are needed to describe and understand socio-economic systems well enough to address practical challenges of humanity (such as the financial and economic crisis) more successfully \cite{FuturIcT}.


\begin{thebibliography}{99}
\bibitem{Comte} A. Comte, {\it Social Physics: From the Positive Philosophy} (Calvin Blanchard, New York, 1856). 
\bibitem{Comte2} A. Comte, {\it Course on Positive Philosophy} (1830-1842).
\bibitem{practical} D. Helbing, Grand socio-economic challenges, Working Paper, ETH Zurich (2010).
\bibitem{Bollinger} L. C. Bollinger, Announcing the Columbia committee on global thought, see http://www.columbia.edu/cu/president/.../051214-committee-global-thought.html
\bibitem{Spencer} H. Spencer, {\it The Principles of Sociology} (Appleton, New York, 1898; the three volumes were originally published in serial form between 1874 and 1896).
\bibitem{Bertalanffy} L. von Bertalanffy, {\it General System Theory: Foundations, Development, Applications} (George Braziller, New York, 1968).
\bibitem{Epstein} J. M. Epstein, {\it Generative Social Science. Studies in Agent-Based Computational Modeling} (Princeton University, 2006), p. 51.
\bibitem{quote} F. Dyson, A meeting with Enrico Fermi. {\it Nature} {\bf 427}, 297 (2004). 
\bibitem{hysteresis} I. D. Mayergoyz, {\it Mathematical Models of Hysteresis and their Applications} (Academic Press, 2003).
\bibitem{phasetrans} H. E. Stanley {\it Introduction to Phase Transitions and Critical Phenomena} (Oxford University, 1987).
\bibitem{Zeeman} E. C. Zeeman, ed. {\it Catastrophe Theory} (Addison-Wesley, London, 1977).
\bibitem{chaos} H. G. Schuster and W. Just, {\it Deterministic Chaos} (Wiley-VCH, Weinheim, 2005).
\bibitem{noiseind} W. Horsthemke and R. Lefever, {\it Noise-Induced Transitions: Theory and Applications in Physics, Chemistry, and Biology} (Springer, 1983).  
\bibitem{Kiss} ``KISS principle'' at Wikipedia.org, see http://en.wikipedia.org/wiki/KISS\_principle
\bibitem{Einstein} A. Einstein, ``On the Method of Theoretical Physics''. The Herbert Spencer Lecture, delivered at Oxford (10 June 1933); also published in {\it Philosophy of Science} {\bf 1}(2), p. 165 (April 1934). %, pp. 163-169., p. 165.
\bibitem{Box} G. E. P. Box and N. R. Draper, {\it Empirical Model-Building and Response Surfaces} (Wiley, 1987), pp. 74+424.
\bibitem{EPJB1} D. Helbing, Derivation of non-local macroscopic traffic equations and consistent traffic pressures from microscopic car-following models. {\it European Physical Journal B} {\bf 69}(4), 539-548 (2009), see also http://www.soms.ethz.ch/research/traffictheory
\bibitem{RMP} D. Helbing, Traffic and related self-driven many-particle systems. {\it Reviews of Modern Physics} {\bf  73}, 1067-1141 (2001). 
\bibitem{Wisdom} F. Galton, Vox populi. {\it Nature} {\bf 75}, 450-451 (1907). 
\bibitem{analytical} D. Helbing {\it et al.}, see collection of publications on analytical traffic flow theory at http://www.soms.ethz.ch/research/traffictheory 
\bibitem{Epstein2} J. M. Epstein, Why model? {\it Journal of Artificial Societies and Social Simulation} {\bf 11}(4), 12 (2008), see http://jasss.soc.surrey.ac.uk/11/4/12.html
\bibitem{turbulence} P. A. Davidson, {\it Turbulence} (Cambridge University, Cambridge, 2004).
\bibitem{Weidlich} W. Weidlich, {\it Sociodynamics: A Systemic Approach to Mathematical Modelling in the Social Sciences} (Dover, 2006). 
\bibitem{exp1} J. H. Kagel and A. E. Roth, {\it The Handbook of Experimental
Economics} (Princeton University, Princeton, NJ, 1995).
\bibitem{exp2} F. Guala, {\it The Methodology of Experimental Economics} (Cambridge
University Press, 2005).
\bibitem{FutureExperimenting} D. Helbing and W. Yu (2010) The future of social experimenting. {\it Proceedings of the National Academy of Sciences USA (PNAS)} {\bf 107}(12), 5265-5266; see also http://www.soms.ethz.ch/research/socialexperimenting
\bibitem{datamining} O. Maimon and L. Rokach, {\it The Data Mining and Knowledge Discovery Handbook} (Springer, 2005).
\bibitem{networks} M. O. Jackson, {\it Social and Economic Networks} (Princeton University, 2008).
\bibitem{agentbased} N. Gilbert (ed.) {\it Computational Social Science} (Sage, 2010).
\bibitem{Schweitzer} F. Schweitzer (ed.) {\it Self-Organization of Complex Structures: From Individual to Collective Dynamics} (CRC, 1997).
\bibitem{complex} J. H. Miller and S. E. Page, {\it Complex Adaptive Systems: An Introduction to Computational Models of Social Life}
(Princeton University, Princeton, NJ, 2007).
\bibitem{criticalphen} D. Sornette, {\it Critical Phenomena in Natural Sciences. Chaos, Fractals, Selforganization and Disorder: Concepts and Tools} (Springer, Berlin, 2006). 
\bibitem{extreme} S. Albeverio, V. Jentsch, and H. Kantz (eds.), {\it Extreme Events in Nature and Society} (Springer, Berlin, 2005).
\bibitem{AI} D. Floreano and C. Mattiussi, Bio-Inspired Artificial Intelligence: Theories, Methods, and Technologies (MIT, Cambridge, MA, 2008).
\bibitem{robotics} S. Nolfi and D. Floreano, {\it Evolutionary Robotics : The Biology, Intelligence, and Technology of Self-Organizing Machines} (MIT, Cambridge, MA, 2000).
\bibitem{BioLogistics} D. Helbing, A. Deutsch, S. Diez, K. Peters, Y. Kalaidzidis, K. Padberg, S. L\"ammer, A. Johansson, G. Breier, F. Schulze, and M. Zerial, Biologistics and the struggle for efficiency: Concepts and perspectives. {\it Advances in Complex Systems} {\bf 12}(6), 533-548 (2009). 
\bibitem{SystemicInstability} D. Helbing, System risks in society and economics. Sante Fe Institute Working Paper \#09-12-044 (2009), see http://www.santafe.edu/media/workingpapers/09-12-044.pdf
\bibitem{TopiCS} M. Moussaid, S. Garnier, G. Theraulaz, and D. Helbing, Collective information processing and pattern formation
in swarms, flocks, and crowds. {\it Topics in Cognitive Science} {\bf 1}(3), 469-497 (2009).
\bibitem{PLoS} D. Helbing, A. Szolnoki, M. Perc, and G. Szab\'o, Evolutionary establishment of moral and double moral standards through spatial interactions. {\it PLoS Computational Biology} {\bf 6}(4), e1000758 (2010). 
\bibitem{Popper} K. R. Popper, {\it The Logic of Scientific Discovery} (Hutchinson, 1959); original German version: {\it Logik der Forschung} (Springer, Vienna, 1935).
\bibitem{inLevin} D. Tilman, D. Wedin, and J. Knops, Productivity and sustainability influenced
by biodiversity in grassland ecosystems. {\it Nature} {\bf 379}, 718-720 (1996).
\bibitem{TraulsenExp} A. Traulsen, D. Semmann, R. D. Sommerfeld, H.-J. Krambeck, and M. Milinski, Human strategy updating in evolutionary games. {\it Proceedings of the National Academy of Sciences USA (PNAS)} {\bf 107}(7), 2962-2966 (2010).
\bibitem{Wagner} E. Brockfeld, R. D. K\"uhne, and P. Wagner, Calibration and validation of microscopic traffic flow models. {\it Transportation Research Board} {\bf 1876}, 62-70 (2004).
\bibitem{variable} A. Kesting and M. Treiber, Calibrating car-following models by using trajectory data: Methodological study. {\it Transportation Research Record} {\bf 2088}, 148-156 (2008).
\bibitem{Goedel} K. G\"odel, {\it On Formally Undecidable Propositions of Principia Mathematica and Related Systems} (Basic, New York, 1962).
\bibitem{WisdomExp} J. Lorenz, H. Rauhut, F. Schweitzer, and D. Helbing, How social influence undermines the wisdom of crowds. Submitted (2010).
\bibitem{Planck} Max Planck: ``An important scientific innovation rarely makes its way by gradually winning over and converting its opponents,  but rather because its opponents eventually die, and a new generation grows up that is familiar with it.''
\bibitem{Kuhn} T. S. Kuhn, {\it The Structure of Scientific Revolutions} (University of Chicago, 1962).
\bibitem{Vicsek} T. Vicsek, The bigger picture. {\it Nature} {\bf 418}, 131 (2002).
\bibitem{Pietronero} L. Pietronero, Complexity ideas from condensed matter and statistical physics. {\it europhysicsnews} {\bf 39}(6), 26-29.
\bibitem{Mikhailov} A. S. Mikhailov and V. Calenbuhr, {\it From Cells to Societies. Models of Complex Coherent Action} (Springer, Berlin, 2002).
\bibitem{Omerod} R. A. Bentley and P. Omerod, Agents, intelligence, and social atoms. Preprint available at http://www.paulormerod.com/pdf/Bentley\_OrmerodSept09.pdf
\bibitem{Gintis} H. Gintis, {\it The Bounds of Reason: Game Theory and the Unification of the Behavioral Sciences} (Princeton University, 2009).
\bibitem{Omerod2} P. Omerod, What can agents know? The feasibility of advanced cognition in social and economic systems. {\it  Communication, Interaction and Social Intelligence} (2008), see http://www.paulormerod.com/pdf/AISB08\%20Whatcanagentsknow\%20Paul\%20Ormerod.pdf
\bibitem{Fehr} E. Fehr and K. M. Schmidt,  A theory of fairness, competition, and cooperation. {\it The Quarterly Journal of Economics} {\bf 114}(3), 817-868 (1999).
\bibitem{Frey} B. Frey, {\it Economics as a Science of Human Behaviour: Towards a New Social Science Paradigm} (Kluwer Academics, Dordrecht, 1999).
\bibitem{prospect} D. Kahneman and A. Tversky, Prospect theory: An analysis of decision under risk. {\it Econometrica} {\bf 47}(2), 263-291 (1979).
\bibitem{satisficing} H. A. Simon, A behavioral model of rational choice. {\it The Quarterly Journal of Economics} {\bf 69}(1), 99-118 (1955).
\bibitem{Gigerenzer} G. Gigerenzer, P. M. Todd, and the ABC Research Group, {\it Simple Heuristics That Make Us Smart}  (Oxford University, 2000).
\bibitem{McFadden} D. McFadden, Conditional logit analysis of qualitative choice behaviour,
in P. Zarembka (ed.) {\it Frontiers of Econometrics} (Academic Press, New York, 1974), pp. 105-142.
\bibitem{QuantSoc} D. Helbing, {\it Quantitative Sociodynamics. Stochastic Methods and Models of Social Interaction Processes} (Kluwer Academic, Dordrecht, 1995).
\bibitem{particle1} P. Woit, {\it Not Even Wrong: The Failure of String Theory and the Search for Unity in Physical Law for Unity in Physical Law} (Basic, New York, 2006).
\bibitem{particle2} L. Smolin, {\it The Trouble With Physics: The Rise of String Theory, The Fall of a Science, and What Comes Next} (Mariner, Boston, 2007).
\bibitem{pluralistic} J. Rotmans and M. B. A. van Asselt, Uncertainty management in integrated assessment modeling: Towards a pluralistic approach. {\it Environmental Monitoring and Assessment} {\bf 69}(2), 101-130 (2001).  
\bibitem{possibilistic} D. Dubois and H. Prade, Possibilistic logic: a retrospective and prospective view. {\it Fuzzy Sets and Systems}
{\bf 144}(1), 3-23 (2004). 
\bibitem{climate} V. Lucarini, Towards a definition of climate science. {\it Int. J. Environment and Pollution} {\bf 18}(5), 413-422 (2002).
\bibitem{Vespignani} A. Vespignani, Predicting the behavior of techno-social systems. {\it Science} {\bf 325}, 425-428 (2009).
\bibitem{FuturIcT} D. Helbing, The FuturIcT knowledge accelerator: Unleashing the power of information for a sustainable future, Project Proposal (2010), see http://arxiv.org/abs/1004.4969 and http://www.futurict.eu

\end{thebibliography}
\end{document}